\documentclass{article}

\usepackage{arxiv}
\usepackage{amsmath}
\usepackage{amssymb}
\usepackage{mathtools,comment}
\usepackage{natbib}
\usepackage[utf8]{inputenc} % allow utf-8 input
\usepackage[T1]{fontenc}    % use 8-bit T1 fonts
\usepackage{hyperref}       % hyperlinks
\usepackage{url}            % simple URL typesetting
\usepackage{booktabs}       % professional-quality tables
\usepackage{amsfonts}       % blackboard math symbols
\usepackage{nicefrac}       % compact symbols for 1/2, etc.
\usepackage{microtype}      % microtypography
\usepackage{lipsum}
\usepackage{graphicx}
\graphicspath{ {./images/} }
\newtheorem{theorem}{Theorem}

\title{Sufficient conditions for proper posteriors in fully-Bayesian Functional PCA}

\author{
 Joseph Sartini \\
  Department of Biostatistics\\
  Johns Hopkins University\\
  Baltimore, MD 21205 \\
  \texttt{jsartin1@jhu.edu} \\
   \And
 Scott Zeger \\
  Department of Biostatistics\\
  Johns Hopkins University\\
  Baltimore, MD 21205 \\
  \And
 Ciprian Crainiceanu \\
  Department of Biostatistics\\
  Johns Hopkins University\\
  Baltimore, MD 21205 \\
}

\begin{document}
\maketitle
\begin{abstract}
    In a fully-Bayesian Functional Principal Components Analysis (FPCA) the principal components are treated as unknown infinite-dimensional parameters. By projecting the functional principal components on a rich orthonormal spline basis, we show that orthonormality of the principal components is equivalent to orthonormality of the spline coefficients. A penalty on the integral of the second derivative of the functional principal components can be induced on the spline coefficients, where each function has its own smoothing parameter. Finally, each smoothing parameter is treated as an inverse variance component in the associated mixed effects model. In this work, we demonstrate that no additional conditions are required to ensure that the corresponding smoothing prior, and thus the posterior distribution, is proper. This allows the choice of less informative priors, such that smoothing is driven by the data.
\end{abstract}

\section{Background}\label{sec:background}

Functional principal components analysis (FPCA) \cite{ramsaysilv2005, crainiceanu2024book} is a popular data analytic method. The FPCA model assumes that the observed  data takes the form $W_i(t)= X_i(t) + \epsilon_i(t)$ for $i = 1, \ldots, N$ and $t\in[0,1]$, where $X_i(t)$ are realizations of a zero mean $L_2[0,1]$-integrable latent process, $X_i(t)$ and $\epsilon_{i}(t)$ are mutually uncorrelated, and $\epsilon_i(t)$ are uncorrelated errors with homogeneous variance $\sigma^2_\epsilon$.  Letting $K_X(\cdot,\cdot)$ denote the covariance operator of $X_i(\cdot)$, the Kosambi-Karhunen-Loève (KKL) theorem \citep{kosambi_statistics_1943, karhunen_uber_1947, loeve_probability_1978} provides the decomposition
\begin{equation}
    X_i(t)=\sum_{k = 1}^\infty \xi_{ik} \phi_k(t)\;,
\end{equation}
where: (1) $\phi_k(t)$ are the orthonormal basis in $L_2[0,1]$ corresponding to the eigenfunctions of $K_X(\cdot,\cdot)$ and (2) the scores $\xi_{ik}$ have zero mean, are uncorrelated, and have variances ${\rm Var}(\xi_{ik})=\lambda_k$, where  $\lambda_1\geq\lambda_2\geq \ldots\geq 0$ are the eigenvalues of $K_X(\cdot,\cdot)$. When $\lambda_k$ converges quickly to zero, the model can be approximated by
\begin{equation}\label{eqn:FPCA}
X_i(t) \approx \sum_{k = 1}^K \xi_{ik} \phi_k(t) + \epsilon_i(t)\;,
\end{equation}
where $K$ is a constant beyond which $\sum_{k=K+1}^\infty \lambda_k$ is negligible and $\epsilon_i(t)$ are once again uncorrelated errors with the same variance $\sigma^2_\epsilon$.

A large and active literature is dedicated to fitting model~\eqref{eqn:FPCA} under the assumption that $X_i(t)$ is a Gaussian process, which is equivalent to the assumptions that $\xi_{ik}\sim N(0,\lambda_k)$, $\epsilon_i(t)\sim N(0,\sigma^2_\epsilon)$, and $\xi_{ik}$ and $\epsilon_i(t)$ are mutually independent. Most approaches obtain an estimator $\widehat{K}_X(\cdot,\cdot)$ of the covariance operator $K_X(\cdot,\cdot)$, obtain eigenfunction estimates  $\widehat{\phi}_k(\cdot)$ of $\phi_k(\cdot)$ by diagonalizing the estimated covariance $\widehat{K}_X(\cdot,\cdot)$, and then treat $\widehat{\phi}_k(\cdot)$ as fixed in subsequent analyses. 

\begin{comment}
In a Bayesian context, priors on all variance components are standard, and here we will use inverse Gamma priors as discussed in \citep{crainiceanuwinbugs,crainiceanu_bayesian_2010,jiang2025}.
\end{comment}

In a series of recent papers \citep{Sartini11022026,sartini2025bayesianmultivariatesparsefunctional}, we propose a fully-Bayesian FPCA approach that treats the functions $\phi_k(t)$, $k=1,\ldots,K$ as unknown parameters and obtains the full joint distribution of all model parameters given the observed data. A key component of this approach is the spline expansion of the infinite dimensional functions $\phi_k(t)$ as $\phi_k(t) = \mathbf{B}(t)\psi_k$ for each $k = 1,\ldots, K$, where  $\mathbf{B}(t) = \{B_1(t), \ldots, B_Q(t)\}$ is a set of $Q$ orthonormal spline basis functions. The basis dimension $Q$ is set large enough to capture the complexity of the first $K$ eigenfunctions and is inherently constrained such that $Q \geq K$. This spline expansion effectively replaces the infinite dimensional functions $\phi_k(\cdot)\in L_2[0,1]$ with the $Q$-dimensional vectors $\psi_k$, and it can be shown that the $\phi_k(t)$ are orthonormal in $L_2[0,1]$ if and only if the vectors $\psi_k$ are orthonormal in $\mathbb{R}^Q$. Indeed, if $\boldsymbol{\Psi} = [\psi_1|\ldots|\psi_K]$ is the $Q\times K$ dimensional matrix obtained by binding the $Q\times 1$ dimensional vectors of spline coefficients $\psi_k$, then the  eigenfunctions $\phi_k(\cdot)$ are orthonormal if and only if $\boldsymbol{\Psi}^\top\boldsymbol{\Psi}=\mathbf{I}_Q$, the identity matrix of dimension $Q$. This, by definition, is equivalent to  $\boldsymbol{\Psi}\in \mathcal{V}_{K,Q}$, where $\mathcal{V}_{K,Q}$ is the $(K,Q)$-Stiefel manifold \citep{james1976}, or the space of $Q\times K$ matrices with orthonormal columns. 

Using this spline expansion, we can induce priors on the $\phi_k(t)$ through the orthonormal vectors $\psi_k$. To be precise, we can induce smoothness on the eigenfunctions using the well-known penalty on the integral of the square of the second derivative introduced by Grace Wahba \citep{wahba_1990, speckman_fully_2003}. Other penalties are possible, but the Wahba prior has well-studied and favorable properties. Note that $\phi_k''(t)=\mathbf{B}''(t)\psi_k$, where $\mathbf{B}''(t)$ is the $1 \times Q$-dimensional vector with the $q$-th entry equal to $B''_q(t)$, the second derivative of $B_q(\cdot)$ evaluated at $t$. Therefore, $\{\phi''_k(t)\}^2 = \psi_k^\top \{\mathbf{B}''(t)\}^\top \mathbf{B}''(t) \psi_k$ and $\int_0^1 \{\phi''_k(t)\}^2 dt = \psi_k^\top \mathbf{P}\psi_k$, where $\mathbf{P}$ is a $Q \times Q$ dimensional matrix with the $(p,q)$ entry equal to $\int_0^1 B_p''(t)B_q''(t)dt$.  For Wahba's original integrated squared second derivative penalty, functions of polynomial order less than two are not penalized, resulting in potentially singular $\mathbf{P}$ depending on basis choice $\mathbf{B}(t)$.

Adding the Wahba prior to each eigenfunction is equivalent to applying the 
(possibly degenerate) normal smoothing priors $h_k^{R/2} \exp(-h_k \psi_k^\top \mathbf{P} \psi_k/2)$, where $h_k$ is the smoothing parameter corresponding to eigenfunction $k$ and $R\leq Q$ is the rank of $\mathbf{P}$. A closer look at this prior reveals that $h_k \mathbf{P}$ can be viewed as the precision matrix for a multivariate normal with zero-mean, where the smoothing parameter $h_k$ is an unknown precision \citep{ruppert2003semiparametric,wood2017gam}. Combining a uniform prior on $\boldsymbol{\Psi}$ over the Stiefel manifold, which enforces orthonormality of the $\phi_k(t)$, with a collection of Wahba priors on the eigenfunctions produces the conditional prior on $\boldsymbol{\Psi}$ given $\mathbf{H}$. This prior, up to a constant, is the following:
\begin{equation}
p(\boldsymbol{\Psi}|\mathbf{H}) = \left\{\prod_{k=1}^K h_k^{R/2} \exp(-h_k \psi_k^\top \mathbf{P} \psi_k/2)\right\} \times \mathbb{I}(\boldsymbol{\Psi} \in \mathcal{V}_{K,Q})\;,
\label{eqn:cond_prior}
\end{equation}

where $\mathbf{H} = \{h_1, \ldots h_K\}$ is used to represent the set of smoothing parameters and $\mathbb{I}(\cdot)$ is the indicator function. Each smoothing parameter $h_k$ controls the complexity of one eigenfunction $\phi_k(t)$. This differential smoothing is necessary, as the complexity of $\phi_k(t)$ tends to increase with $k$.

As recommended by \cite{crainiceanuwinbugs,crainiceanu_bayesian_2010,jiang2025}, we use independent Gamma priors on $h_k$. This leads to the following prior on the smoothing parameters: $p(\mathbf{H})=\prod_{k = 1}^K G(h_k|\alpha_\psi, \beta_\psi)$, where $G(x|a,b)$ denotes the Gamma distribution with shape $a$ and rate $b$ evaluated at $x$. We aim to choose hyperparameters $\alpha_\psi$ and $\beta_\psi$ such that the prior is weakly informative. Therefore, the proposed prior is $p(\mathbf{\Psi}, \mathbf{H})  = p(\mathbf{\Psi}|\mathbf{H}) p(\mathbf{H})$, which has the following explicit form up to a normalizing constant:
\begin{equation} \label{eqn:prior_form}
 p(\boldsymbol{\Psi}, \mathbf{H})  =
 \left(\prod_{k = 1}^K h_k^{R/2} \right)\times \exp\left(-\frac{1}{2}\sum_{k = 1}^K h_k \psi_k^\top \mathbf{P} \psi_k\right) \times \mathbb{I}(\boldsymbol{\Psi} \in \mathcal{V}_{K,Q}) \times \prod_{k = 1}^K G(h_k|\alpha_\psi, \beta_\psi)\;,
\end{equation}

This distribution is not known, so we must find sufficient conditions on $\alpha_\psi$, $\beta_\psi$ to ensure that $p(\boldsymbol{\Psi}, \mathbf{H})$ is a proper distribution, that is the integral $\int\int p(\boldsymbol{\Psi}, \mathbf{H})d\boldsymbol{\Psi}d\mathbf{H}<\infty$. 

Given the well-defined Gaussian likelihood of FPCA, proper prior distributions ensure the model is well-specified and the posterior integrates. While it is possible for the posterior to integrate even for an improper choice of prior, analytic evaluation of the posterior for fully-Bayesian FPCA has proven prohibitive.

\section{Main result: Sufficient conditions to ensure the prior is proper}\label{sec:main_result}

\begin{theorem}\label{thm:main_thm}
The joint prior $p(\boldsymbol{\Psi}, \mathbf{H})$ is proper for all valid hyperparameter values of $\alpha_\psi > 0$ and $\beta_\psi > 0$.
\end{theorem}

This result is practical, because it facilitates choosing $\alpha_\psi$ and $\beta_\psi$ such that the prior on each $h_k$ is not informative, ensuring that the smoothing parameter posterior distributions are driven by the data.

We can assess how informative various choices of $\alpha_\psi, \beta_\psi$ are by examining the form of the conditional posterior of smoothing parameter $h_k$ given the other parameters and the data \citep{Sartini11022026}. We denote this quantity as $[h_k|{\rm others}]$ and continue to use $R$ to refer to the rank of the penalty $\mathbf{P}$.
\begin{equation}\label{eqn:sp_cond}
    [h_k|{\rm others}] \propto h_k^{(\alpha_\psi + R/2) - 1} \exp\left\{-\left(\beta_\psi + \frac{\psi_k^\top \mathbf{P}\psi_k}{2} \right)h_k\right\}
\end{equation}

One can recognize the form of the Gamma distribution in Equation~\ref{eqn:sp_cond}, such that $h_k$ has the following conditional posterior distribution given the other parameters and data:
\begin{equation}\label{eqn:sp_cond_full}
 h_k|{\rm others} \sim G(h_k|\alpha_\psi + R/2, \beta_\psi + \psi_k^\top \mathbf{P}\psi_k/2)
\end{equation}

Based upon this conditional distribution, one finds that choosing $\alpha_\psi << R/2$ and $\beta_\psi << \psi_k^\top \mathbf{P}\psi_k/2$ will provide an uninformative prior. 

As increasing $\beta_\psi$ may be computationally beneficial due to the reduced prior sparsity, it is important to understand exactly how this parameter impacts the informativeness of the prior. The normality of the $\psi_k$ vectors, such that $\psi_k^\top \psi_k = 1$, implies $\psi_k^\top \mathbf{P}\psi_k/2$ has the form of a Rayleigh quotient. This quantity can be described as a linear combination of the eigenvalues of the matrix $\mathbf{P}/2$, providing an idea of the data contribution to the rate parameter of each $h_k$ conditional posterior. Alternatively, one can empirically assess the posterior scale of each $\psi_k^\top \mathbf{P} \psi_k/2$ when using a very small $\beta_\psi$. 

\section{Proof}\label{sec:propriety}
The goal is to show that the integral $\int \int p(\boldsymbol{\Psi},\mathbf{H})d\boldsymbol{\Psi}d\mathbf{H}<\infty$. We use the following strategy:
\begin{itemize}
\item[1.]  Show that $\int\int p(\boldsymbol{\Psi},\mathbf{H})d\boldsymbol{\Psi}d\mathbf{H}=\int\{\int p(\boldsymbol{\Psi}|\mathbf{H})p(\mathbf{H})d\mathbf{H}\}d\boldsymbol{\Psi}$ and integrate over the set of smoothing parameters $\mathbf{H} = \{h_1, \ldots, h_K\}$.
\item[2.] Simplify and place an upper bound on $\int p(\boldsymbol{\Psi}|\mathbf{H})p(\mathbf{H})d\mathbf{H}$.
\item[3.] Using monotonicity, integrate over the Stiefel manifold $\mathcal{V}_{K,Q}$ to find a finite upper bound on the integral of interest, $\int\int p(\boldsymbol{\Psi},\mathbf{H})d\boldsymbol{\Psi}d\mathbf{H}$

\end{itemize}

\subsection{Part 1: Integration over $\mathbf{H}$}\label{subsec:int_H}

As $p(\boldsymbol{\Psi}, \mathbf{H})\geq 0$, we have $\int \int p(\boldsymbol{\Psi}, \mathbf{H})d\boldsymbol{\Psi}d\mathbf{H} = \int \{\int p(\boldsymbol{\Psi}|\mathbf{H})p(\mathbf{H})d\mathbf{H}\}d\boldsymbol{\Psi}$, where the outer integral is over the Stiefel manifold $\mathcal{V}_{K,Q}$. We first focus on the interior integral over the set of non-negative smoothing parameters $\mathbf{H} = \{h_1, \ldots, h_K\}$. We can substitute in the known forms of the conditional $p(\boldsymbol{\Psi}|\mathbf{H})$ and prior $p(\mathbf{H})$.
\begin{align}
\int p(\boldsymbol{\Psi}|\mathbf{H})p(\mathbf{H})d\mathbf{H} & = \int \exp\left(-\frac{1}{2}\sum_{k = 1}^K h_k \psi_k^\top \mathbf{P} \psi_k\right) \prod_{k = 1}^K h_k^{R/2}G(h_k|\alpha_\psi, \beta_\psi) d\mathbf{H} \label{eqn:int_H} \\
& = \int \prod_{k = 1}^K \frac{\beta_\psi^{\alpha_\psi}}{\Gamma(\alpha_\psi)} h_k^{(\alpha_\psi + R/2) - 1}\exp\left(-\left[\beta_\psi + \psi_k^\top \mathbf{P} \psi_k/2\right]h_k\right)d \mathbf{H}\label{eqn:int_H_2} 
\end{align}

We recognize in Equation~\ref{eqn:int_H_2} that the integrand factors into components associated with each smoothing parameter $h_k$. As such, integrating over the set $\mathbf{H}$ can be accomplished by integrating over each $h_k$ separately.  Further, we can recognize that each such component has the form of a gamma distribution. We can use these two insights together to evaluate the integral. Here, we let $NC(\alpha, \beta)$ represent the normalizing constant for the gamma distribution with shape $\alpha$ and rate $\beta$.
\begin{align}
    \int p(\boldsymbol{\Psi}|\mathbf{H})p(\mathbf{H})d\mathbf{H} & = \prod_{k = 1}^K \int_0^{\infty} \frac{\beta_\psi^{\alpha_\psi}}{\Gamma(\alpha_\psi)} h_k^{(\alpha_\psi + R/2) - 1}\exp\left(-\left[\beta_\psi + \psi_k^\top \mathbf{P} \psi_k/2\right]h_k\right)d h_k\\
    & = \prod_{k = 1}^K \frac{\beta_\psi^{\alpha_\psi}}{\Gamma(\alpha_\psi)} \int_0^\infty h_k^{(\alpha_\psi + R/2) - 1}\exp\left(-\left[\beta_\psi + \psi_k^\top \mathbf{P} \psi_k/2\right]h_k\right)d h_k\\
    & = \prod_{k = 1}^K \frac{\beta_\psi^{\alpha_\psi}}{\Gamma(\alpha_\psi)} NC(\alpha_\psi + R/2, \beta_\psi + \psi_k^\top \mathbf{P} \psi_k/2)
\end{align}

Plugging in the known form of the gamma normalizing constant yields the following.
\begin{align}
& \int p(\boldsymbol{\Psi}|\mathbf{H})p(\mathbf{H})d\mathbf{H} \\
& = \prod_{k = 1}^K \frac{\beta_\psi^{\alpha_\psi}}{\Gamma(\alpha_\psi)} \times \frac{\Gamma(\alpha_\psi + R/2)}{(\beta_\psi + \psi_k^\top \mathbf{P}\psi_k/2)^{\alpha_\psi + R/2}}\label{eqn:raw_intH}
\end{align}

The result is now a function only of fixed hyper-parameters and the orthonormal eigenfunction spline coefficients $\psi_k$.

\subsection{Part 2: Simplifying and bounding the integral over $\mathbf{H}$}\label{subsec:sim_bound}

We begin simplification of Equation~\ref{eqn:raw_intH} by separating the constant elements and those which depend on the eigenfunction spline coefficients $\psi_k$.
\begin{align}
\int p(\boldsymbol{\Psi}|\mathbf{H})p(\mathbf{H})d\mathbf{H} & = \prod_{k = 1}^K \frac{\beta_\psi^{\alpha_\psi}}{\Gamma(\alpha_\psi)} \times \frac{\Gamma(\alpha_\psi + R/2)}{\beta_\psi^{\alpha_\psi + R/2}(1 + \psi_k^\top\mathbf{P}\psi_k/\{2\beta_\psi\})^{\alpha_\psi + R/2}}\\
& = \prod_{k = 1}^K \frac{\Gamma(\alpha_\psi + R/2)}{\Gamma(\alpha_\psi)\beta_\psi^{R/2}} \times (1 + \psi_k^\top\mathbf{P}\psi_k/\{2\beta_\psi\})^{-(\alpha_\psi + R/2)}\\
& = \left\{\frac{\Gamma(\alpha_\psi + R/2)}{\Gamma(\alpha_\psi)\beta_\psi^{R/2}}\right\}^K \times \prod_{k = 1}^K \left(1 + \psi_k^\top \left\{\frac{\mathbf{P}}{2\beta_\psi}\right\}\psi_k\right)^{-(\alpha_\psi + R/2)}
\end{align}

As $\alpha_\psi$ and $\beta_\psi$ are positive and finite, the first term $\left\{\frac{\Gamma(\alpha_\psi + R/2)}{\Gamma(\alpha_\psi)\beta_\psi^{R/2}}\right\}^K$ will be a finite constant with respect to the eigenfunction spline coefficients $\psi_k$. 

Turn attention now to the quadratic terms $\psi_k^\top \left\{\frac{\mathbf{P}}{2\beta_\psi}\right\} \psi_k$. As the $\psi_k$ are normal by construction, such that $\psi_k^\top \psi_k = 1$ for all $k = 1, \ldots, K$, each term $\psi_k^\top \left\{\frac{\mathbf{P}}{2\beta_\psi}\right\} \psi_k$ has the form of a Rayleigh quotient. We can therefore invoke the following inequalities for each $k$, where $\lambda_{\text{max}}(\mathbf{A})$ and $\lambda_{\text{min}}(\mathbf{A})$ denote the largest and smallest eigenvalues of the generic argument $\mathbf{A}$ (see Theorem 4.2.2 in \cite{Horn_Johnson_2012}).
\begin{equation}
\lambda_{\text{max}}\left(\frac{\mathbf{P}}{2\beta_\psi}\right) \geq \psi_k^\top \left\{\frac{\mathbf{P}}{2\beta_\psi}\right\} \psi_k \geq \lambda_{\text{min}}\left(\frac{\mathbf{P}}{2\beta_\psi}\right)\label{eqn:rayleigh}
\end{equation}

As the eigenvalues of $c\mathbf{A}$ for scalar $c$ and matrix $\mathbf{A}$ are precisely $c\lambda_i$, where $\lambda_i$ are the eigenvalues of $\mathbf{A}$, the bounds in Equation~\ref{eqn:rayleigh} simplify as follows.
\begin{equation}\label{eqn:rayleigh_2}
\frac{\lambda_{\text{max}}(\mathbf{P})}{2\beta_\psi}\geq \psi_k^\top \left\{\frac{\mathbf{P}}{2\beta_\psi}\right\} \psi_k \geq \frac{\lambda_{\text{min}}(\mathbf{P})}{2\beta_\psi}
\end{equation}

Considering now each factor $\left(1 + \psi_k^\top \left\{\frac{\mathbf{P}}{2\beta_\psi}\right\}\psi_k\right)^{-(\alpha_\psi + R/2)}$. The negative exponent implies that each such term is a decreasing function in the value of the quadratic form $\psi_k^\top \left\{\frac{\mathbf{P}}{2\beta_\psi}\right\}\psi_k$. We can combine this fact with the inequality in Equation~\ref{eqn:rayleigh_2} to find upper bounds for each $k = 1,\ldots, K$.

\begin{equation}
    \left(1 + \psi_k^\top \left\{\frac{\mathbf{P}}{2\beta_\psi}\right\}\psi_k\right)^{-(\alpha_\psi + R/2)} \leq \left(1 + \frac{\lambda_{\text{min}}(\mathbf{P})}{2\beta_\psi}\right)^{-(\alpha_\psi + R/2)}
\end{equation}

Recall now that $\mathbf{P}$ is positive semi-definite, such that $\lambda_{\text{min}}(\mathbf{P}) \geq 0$. We can use this fact to produce a simpler upper bound as follows.
\begin{equation}
    \left(1 + \psi_k^\top \left\{\frac{\mathbf{P}}{2\beta_\psi}\right\}\psi_k\right)^{-(\alpha_\psi + R/2)} \leq \left(1 + 0\right)^{-(\alpha_\psi + R/2)}= 1
\end{equation}

So, we can bound each term $\left(1 + \psi_k^\top \left\{\frac{\mathbf{P}}{2\beta_\psi}\right\}\psi_k\right)^{-(\alpha_\psi + R/2)}$ from above by $1$. Applying this for each $k = 1, \ldots, K$ yields the following simple upper bound on the integral $\int p(\boldsymbol{\Psi}|\mathbf{H})p(\mathbf{H})d\mathbf{H}$ over the smoothing parameters.
\begin{equation}
    \int p(\boldsymbol{\Psi}|\mathbf{H})p(\mathbf{H})d\mathbf{H} \leq \left\{\frac{\Gamma(\alpha_\psi + R/2)}{\Gamma(\alpha_\psi)\beta_\psi^{R/2}}\right\}^K \times 1^K = \left\{\frac{\Gamma(\alpha_\psi + R/2)}{\Gamma(\alpha_\psi)\beta_\psi^{R/2}}\right\}^K \label{eqn:inner_ineq}
\end{equation}

We have thus placed an upper bound on the integral over the smoothing parameters $\{h_1, \ldots, h_K\}$ which does not depend on the spline coefficients $\psi_k$.

\subsection{Part 3: Integrate the upper bound over $\Psi$}\label{subsec:int_psi}

We now return to the integral of interest, $\int \int p(\boldsymbol{\Psi}, \mathbf{H})d\boldsymbol{\Psi}d\mathbf{H}$. By monotonicity of the integral and the inequality derived in Equation~\ref{eqn:inner_ineq}, we can derive the following.
\begin{align}
     \int \int p(\boldsymbol{\Psi}, \mathbf{H})d\boldsymbol{\Psi}d\mathbf{H} & = \int_{\boldsymbol{\Psi} \in \mathcal{V}_{K,Q}} \left\{\int p(\boldsymbol{\Psi}|\mathbf{H})p(\mathbf{H})d\mathbf{H}\right\}d\boldsymbol{\Psi}\\
     & \leq \int_{\boldsymbol{\Psi} \in \mathcal{V}_{K,Q}} \left\{\frac{\Gamma(\alpha_\psi + R/2)}{\Gamma(\alpha_\psi)\beta_\psi^{R/2}}\right\}^K d\boldsymbol{\Psi} \label{eqn:bounding_int}
\end{align} 

We can now evaluate the integral in Equation~\ref{eqn:bounding_int} directly to find that it is finite. Note first that the integrand in Equation~\ref{eqn:bounding_int} does not contain the spline coefficient matrix $\boldsymbol{\Psi} = [\psi_1|\ldots|\psi_K]$. As such, evaluating Equation~\ref{eqn:bounding_int} amounts to integrating a constant over the compact Stiefel manifold $\mathcal{V}_{K,Q}$, which has finite volume \citep{chikuse_statistics_2003}. We denote this volume using $\text{vol}(\mathcal{V}_{K,Q})$. We can therefore evaluate the integral as follows.
\begin{align}
    \int_{\boldsymbol{\Psi} \in \mathcal{V}_{K,Q}} \left\{\frac{\Gamma(\alpha_\psi + R/2)}{\Gamma(\alpha_\psi)\beta_\psi^{R/2}}\right\}^K d\boldsymbol{\Psi} & = \left\{\frac{\Gamma(\alpha_\psi + R/2)}{\Gamma(\alpha_\psi)\beta_\psi^{R/2}}\right\}^K \int_{\boldsymbol{\Psi} \in \mathcal{V}_{K,Q}} 1 d\boldsymbol{\Psi}\\
    & = \left\{\frac{\Gamma(\alpha_\psi + R/2)}{\Gamma(\alpha_\psi)\beta_\psi^{R/2}}\right\}^K \times \text{vol}(\mathcal{V}_{K,Q})
\end{align}

Now, as both factors $\left\{\frac{\Gamma(\alpha_\psi + R/2)}{\Gamma(\alpha_\psi)\beta_\psi^{R/2}}\right\}^K$ and $\text{vol}(\mathcal{V}_{K,Q})$ are finite, we have found a finite upper bound on the integral of interest $\int \int p(\boldsymbol{\Psi}, \mathbf{H})d\boldsymbol{\Psi}d\mathbf{H}$, concluding proof of Theorem~\ref{thm:main_thm}.

\clearpage

\bibliographystyle{plainnat}

\bibliography{references}

\end{document}